\documentclass[preprint,12pt,authoryear]{elsarticle}

\usepackage{amssymb}
\usepackage[normalem]{ulem}

\journal{iScience}

\usepackage{natbib}
\usepackage{amssymb,amsmath,amstext,amsthm,mathtools}
\usepackage{algorithm}
\usepackage{algorithmicx}
\usepackage{algpseudocode}
\usepackage{url}
\usepackage{hyperref}

\usepackage{xcolor}
\newcommand{\pr}[1]{\mathbb{P}\left\{#1\right\}}

\renewcommand{\l}{\left}
\renewcommand{\r}{\right}

\newcommand{\beginsupplement}{%
        \setcounter{table}{0}
        \renewcommand{\thetable}{S\arabic{table}}%
        \setcounter{figure}{0}
        \renewcommand{\thefigure}{S\arabic{figure}}%
     }

\begin{document}

\begin{frontmatter}



\title{Gene expression noise accelerates the evolution of a biological oscillator}


\author[inst1,inst3,inst4]{Yen Ting Lin}

\affiliation[inst1]{organization={Information Sciences Group (CCS-3), Computer, Computational and Statistical Sciences Division, Los Alamos National Laboratory},
            addressline={MS B256, 1 Bikini Atoll}, 
            city={Los Alamos},
            postcode={87545}, 
            state={New Mexico},
            country={USA},}

\author[inst2,inst3,inst5]{Nicolas E.~Buchler}

\affiliation[inst2]{organization={Department of Molecular Biomedical Sciences, North Carolina State University},
            city={Raleigh},
            postcode={27606}, 
            state={North Carolina},
            country={USA}}
            
\affiliation[inst3]{These authors contributed equally.}
\affiliation[inst4]{Correspondence: yentingl@lanl.gov}
\affiliation[inst5]{Lead Contact: nebuchle@ncsu.edu}

\begin{abstract}
Gene expression is a biochemical process, where stochastic binding and un-binding events naturally generate fluctuations and cell-to-cell variability in gene dynamics.  These fluctuations typically have destructive consequences for proper biological dynamics and function (e.g., loss of timing and synchrony in biological oscillators).  Here, we show that gene expression noise counter-intuitively accelerates the evolution of a biological oscillator and, thus, can impart a benefit to living organisms. We used computer simulations to evolve two mechanistic models of a biological oscillator at different levels of gene expression noise.  We first show that gene expression noise induces oscillatory-like dynamics in regions of parameter space that cannot oscillate in the absence of noise. We then demonstrate that these noise-induced oscillations generate a fitness landscape whose gradient robustly and quickly guides evolution by mutation towards robust and self-sustaining oscillation.  These results suggest that noise can help dynamical systems evolve or learn new behavior by revealing cryptic dynamic phenotypes outside the bifurcation point.  

\end{abstract}






\end{frontmatter}


\newcommand{\figCircuitTrajectory}{1}
\newcommand{\figEvolAlgo}{2}
\newcommand{\figRepressorI}{3}
\newcommand{\figTitrationI}{4}
\newcommand{\figTitrationII}{5}
\newcommand{\figTitrationIII}{6}
\newcommand{\figPSD}{7}
\newcommand{\figPSDtraj}{8}
\newcommand{\figBifurcation}{9}


\section{Introduction} \label{sec:introduction}
Organisms have evolved to survive in a dynamic world, where some environmental changes are predictable (e.g. light-dark cycles that arise from the rotation of the Earth) but also include stochastic elements, such as the weather. The ability to sense, compute, and appropriately respond to such changes can provide an evolutionary advantage to organisms. This is perhaps best exemplified by circadian clocks, which are gene regulatory networks that regulate the physiology and behavior of organisms, and which have evolved to oscillate with a $\sim$24-hour period and to be entrained to light-dark cycles. It is thought that the ability to internalize the 24-hour light-dark cycle via a genetic oscillator provided those organisms with the ability to anticipate and prepare for upcoming light changes, rather than simply respond to them \citep{pittendrigh1993,paranjpe2005}. This anticipation provided a selective advantage to those organisms that evolved genetic circuits that oscillate in resonance with the light-dark cycle. This notion is supported by experiments showing that mutant strains with shorter or longer circadian periods have worse fitness than organisms with a 24-hour circadian clock that resonates with the natural light-dark cycle \citep{ouyang1998,dodd2005,wyse2010,spoelstra2016}. 

The ability to internalize an external environmental signal and use this information to predict changes and improve fitness should be negatively impacted by ``misinformation'' or stochastic noise. This is especially true in living systems, which are composed of cells, the fundamental building block of all organisms.  As recognized by Schr{\"o}dinger in his book ``\emph{What is Life?}'', cells are microscopic systems where the number of molecules per cell are expected to exhibit significant fluctuations around their mean. This biochemical noise affects the ability of genetic circuits to faithfully sense, compute, respond to deterministic external signals. For example, experiments in bacteria, plants, and animals have shown that circadian rhythms in single cells have large variation in amplitude, period, and phase \citep{welsh1995,nagoshi2004,liu2007,gould2018,chew2018,li2020}. Thus, an active area of circadian research is to understand how biological systems have evolved to mitigate the destructive effects of stochastic noise on the 24-hour period. For example, noisy circadian clocks in the brain region known as the superchiasmatic nucleus (the primary circadian pacemaker of the animal brain) synchronize with with one another to maintain a coherent 24-hour period and buffer against the effects of biological noise \citep{Welsh2010}. The same principle can also occur at a molecular level within a single cell, where thousands of molecular clocks (i.e., KaiC hexamers) exchange regulators and components to synchronize the molecular clock population within a single cyanobacterium \citep{vanzon2007}. 

The circadian gene network is a nonlinear system, whose oscillatory dynamics depend sensitively on the biophysical parameters (e.g. protein-DNA, protein-protein interactions). Most parameter space is non-oscillatory, and the circadian gene network evolved through the process of genetic mutation (i.e., randomly modification of biophysical parameters) and selection of fitter cells (i.e., those with oscillatory dynamics that resonate with the light-dark cycle). This raises the following questions:  If most of parameter space is non-oscillatory, how long will evolution blindly walk in parameter space until the genetic network crosses a bifurcation point and starts oscillating?  How will gene expression noise, which negative impacts the timing and fitness of circadian clocks, impact the evolution of a circadian oscillator?  

Here, we show that gene expression noise counter-intuitively accelerates the evolution of a biological oscillator and, thus, imparts a benefit to living organisms. We used computer simulation to evolve two different types of oscillatory genetic networks (repressilator and a titration-based circadian clock). In both cases, gene expression noise induced oscillatory-like dynamics in regions of parameter space that cannot oscillate autonomously in the absence of noise. We show how noise-induced oscillation generates a fitness landscape whose gradient robustly and quickly guides evolution by mutation towards self-sustaining oscillation. Last, we identify three distinct types of mechanisms of noise-induced oscillations and discuss how each mechanism responds to noise. 

\section{Results} \label{sec:results}
\subsection{Oscillatory gene circuits}
We studied two genetic circuits that admit oscillatory dynamics in suitable parameter regimes. The first model, which we refer to as the \emph{repressilator}, is a simplified gene expression model of the classic repressilator circuit \citep{elowitz2000SyntheticOscillatoryNetwork}. The second model, referred to as the \emph{titration oscillator}, is a stylized model capturing the mechanism and dynamics of a family of titration-based circadian clocks \citep{vilarMechanismsNoiseresistanceGenetic2002,francois2005,kim2012,karapetyanRoleDNABinding2015,linEfficientAnalysisStochastic2017}. 

\subsubsection{Repressilator}
The repressilator is a three gene circuit of transcriptional repressors, $X$, $Y$, and $Z$, that suppress each other's synthesis. We denote their concentrations by $x$, $y$, and $z$. The molecules are produced according to a zeroth order synthesis term that depends on the concentration of its upstream regulator:
\begin{subequations}
\begin{align}
    \varnothing \xrightarrow{H(z)}{X}, \\
    \varnothing \xrightarrow{H(x)}{Y}, \\
    \varnothing \xrightarrow{H(y)}{Z}.
\end{align}
\end{subequations}
The rate constants of the synthesis of $X$, $Y$ and $Z$ are cyclically and symmetrically regulated by $Z$, $X$, and $Y$, respectively. We used a Hill function to model the negative regulation
\begin{equation}
    H(\phi) := r_0 + \l(r_1-r_0\r) \frac{\phi^h}{\phi^h + \phi_{\ast}^h},
\end{equation}
parameterized by the fully-repressed rate $r_1$, the unrepressed rate $r_0\ge r_1$, half-maximum concentration $\phi_\ast$, and Hill coefficient $h$. In our simulations below, we fixed $\phi_{\ast}=0.5$ and $h=3$ and treated $r_0$ and $r_1$ as adjustable biophysical parameters driven by an evolutionary process detailed in Sec.~\ref{sec:evolution}. The molecules degrade uniformly according to a first-order decaying process:
\begin{subequations}
\begin{align}
    X \xrightarrow{\delta}{\varnothing}, \\
    Y \xrightarrow{\delta}{\varnothing}, \\
    Z \xrightarrow{\delta}{\varnothing},
\end{align} 
\end{subequations}
with a degradation rate $\delta$. With no loss of generality, we fix $\delta=1$ by choosing an appropriate time unit. Our model is a simplified version of the classic repressilator \citep{elowitz2000SyntheticOscillatoryNetwork} because we ignored the {mRNA} populations and we coarse-grained the two-stage (transcription and then translation) process into one effective ``production'' term.  Figure {\figCircuitTrajectory}A shows a schematic diagram of the regulatory network.

\begin{figure}[ht]
\centering
\includegraphics[width=0.9\linewidth]{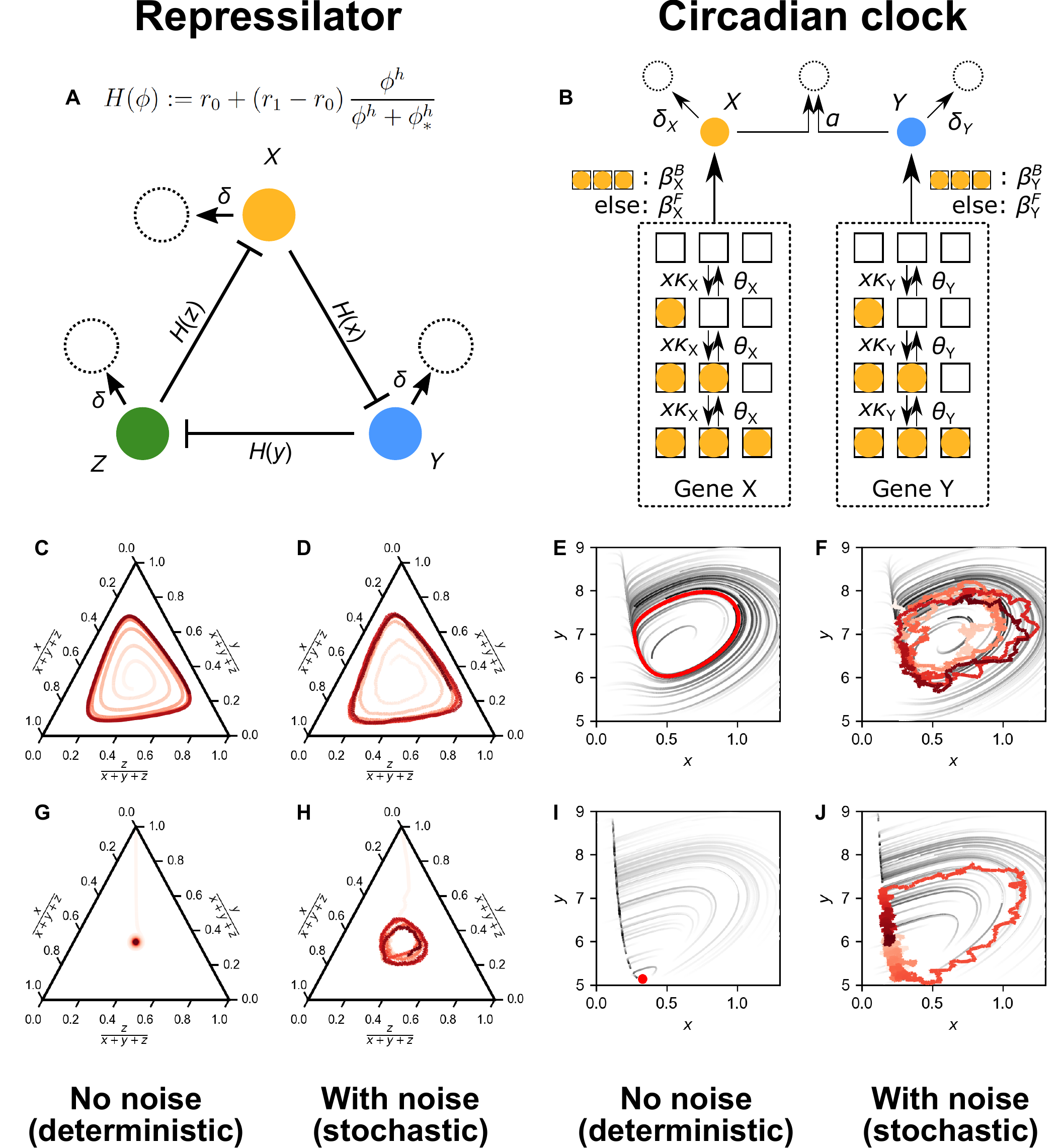}
\caption{{\bf Deterministic and stochastic dynamics of the repressilator and a titration-based circadian clock.} (A) Simplified model of the repressilator, a negative feedback ring oscillator where protein production is given by a repressive Hill function $H(x)$ and proteins are degraded with first-order kinetics. (B) Simplified titration-based circadian clock, where X is an activator that binds both its own promoter (positive feedback) and the promoter of an inhibitor Y. The inhibitor irreversibly binds and sequesters X into an inactive complex (negative feedback loop). (C and D) Deterministic and stochastic simulation of the repressilator for oscillatory parameters, $r_0$=5, $r_1$=0. (E and F) Deterministic and stochastic simulation of the circadian clock for oscillatory parameters, $\beta^F_x$=15, $\beta^B_x$=200. (G and H) Deterministic and stochastic simulation of the repressilator for non-oscillatory parameters, $r_0$=0.2, $r_1$=0. (I and J) Deterministic and stochastic simulation of the circadian clock for oscillatory parameters, $\beta^F_x$=10, $\beta^B_x$=200.}
\label{fig:Figure1}
\end{figure}

\subsubsection{Titration oscillator}

The titration oscillator is a more detailed model that is based on the architecture of the animal and fungal circadian clocks, and explicitly includes transcription factor binding events to circadian promoters.  The titration model is a two gene circuit, which consists of a transcriptional activator $X$ that stimulates its own production (positive feedback) and that of its inhibitor $Y$ (negative feedback). This genetic module is also known as a mixed feedback loop, based on the mixture of positive and negative feedback \citep{francois2005}. 

Both the activator and inhibitor are produced according to zeroth order processes, whose rates depends on a discrete promoter state of their genes, $s_X$ and $s_Y$, both of which take values in $\l\{0,\ldots N\r\}$. We consider the state $s_{X,Y}$ as the number of bound activators on promoter $X$ and $Y$ respectively, and $N$ is the number of binding sites on each promoter.  Based on our previous work, we simplified the titration model in two ways \citep{linEfficientAnalysisStochastic2017}.  First, the activator binds to empty promoter sites sequentially, and the binding and unbinding events between the regulator and promoter sites are modeled as discrete-state Markov processes:
\begin{subequations}
\begin{align}
    s_X \xrightarrow{x \kappa_X } s_X + 1, \\
    s_X \xrightarrow{ \theta_X } s_X - 1, \\
    s_Y \xrightarrow{x \kappa_Y} s_Y + 1, \\
    s_Y \xrightarrow{ \theta_Y } s_Y - 1. 
\end{align}
\end{subequations}

Second, we assumed that maximum transcription ($\beta^B$) only occurs when all the binding sites are ``bound'' by activators. The transcription is lower when one or more binding sites are ``free'' and unbound ($\beta^F$). Thus, the synthesis of the molecules is given by: 

\begin{subequations}
\begin{align}
    \varnothing \xrightarrow{\beta_X\l(s_X\r)}{X}, \\
    \varnothing \xrightarrow{\beta_Y\l(s_Y\r)}{Y}, 
\end{align}
\end{subequations}
where the synthesis rates are
\begin{subequations}
\begin{align}
    \beta_X\l(s_X\r) = \l\{\begin{array}{cl} 
                            \beta_X^F & \text{if } s_X<N, \\
                            \beta_X^B & \text{if } s_X=N, \\
                           \end{array}\r. \\
    \beta_Y\l(s_Y\r) = \l\{\begin{array}{cl} 
                            \beta_Y^F & \text{if } s_Y<N, \\
                            \beta_Y^B & \text{if } s_Y=N, \\
                           \end{array}\r.
\end{align}
\end{subequations}
The molecules degrade according to the following first-order reactions:
\begin{subequations}
\begin{align}
    {X}\xrightarrow{\delta_X}\varnothing , \\
    {Y}\xrightarrow{\delta_Y}\varnothing.
\end{align}
\end{subequations}

\noindent Last, inhibitor will bind and titrate free activator into an inactive complex. This reaction is modeled as an irreversible process:
\begin{equation}
    X+Y \xrightarrow{\alpha} \varnothing. 
\end{equation}
In our simulations below, we fixed $\beta_Y^F=10$, $\beta_Y^B=400$, $\delta_X=1$, $\delta_Y=0.05$, $\kappa_X=60$, $\kappa_Y=45$, $\theta_X = \theta_Y=50$, $N=3$, and $\alpha=10$. We will treat $\beta_X^F$ and $\beta_
X^B$ as adjustable biophysical parameters driven by an evolutionary process detailed in Sec.~\ref{sec:evolution}. Figure {\figCircuitTrajectory}B shows a schematic diagram of the regulatory network.

\subsection{Stochastic simulation of dynamics}

In the above section, we used standard nomenclature of chemical reactions to specify our model. The chemical reactions of the repressilator and titration oscillator were converted into a Chemical Master Equation (see Appendix), which is a formal mathematical representation describing the time-evolution of the joint probability distribution of discrete, molecular species and promoter states undergoing probabilistic chemical reactions. For each cell with a given set of biophysical parameters and initial conditions, we used the Gillespie algorithm \citep{gillespieGeneralMethodNumerically1976,gillespleExactStochasticSimulation1977} to simulate the stochastic time-evolution of discrete molecular species for both the repressilator and titration oscillator CME. An important factor for both the CME and Gillespie simulation is a parameter $\Omega$, which characterizes the cell ``volume'' or size of the biochemical system. Under the well-mixed assumption, the molecular species in a smaller system will exhibit a larger coefficient of variance (i.e., biochemical noise) compared to a larger system. Thus, decreasing or increasing $\Omega$ will increase or decrease the intrinsic biochemical noise in our stochastic Gillespie simulations.

\subsection{Deterministic simulation of dynamics}

The deterministic model is equivalent to the stochastic model in the infinite-population limit, where $\Omega\rightarrow \infty$. For the repressilator CME, a Kramers--Moyal expansion leads to the dynamics of stochastic population densities $\l\{x,y,z\r\}:=N_{\l\{X,Y,Z\r\}}(t)/\Omega$ in the infinite-population limit $\Omega\rightarrow \infty$:
\begin{subequations} \label{eq:ODE-repressilator}
\begin{align} 
    \dot{x} = H\l(y\r) - \delta x, \\
    \dot{y} = H\l(z\r) - \delta y, \\
    \dot{z} = H\l(x\r) - \delta z.
\end{align}
\end{subequations}

For the deterministic titration oscillator model, we invoked an additional fast-switching condition where the binding and unbinding rates ($\kappa$'s and $\theta$'s) are infinitely fast. Under this condition, the discrete promoter state can be represented by a continuous mean-field, quasi-stationary distribution of the continuous variables $x(t)$ and $y(t)$. This adiabatic limit \citep{ackers1982quantitative,hornos2005self,linEfficientAnalysisStochastic2017,hufton2019ClassicalStochasticSystems,hufton2019ModelReductionMethods} leads to simple dynamic equations for the intensive variables $\l\{x,y\r\}:=N_{\l\{X,Y\r\}}(t)/\Omega$
\begin{subequations} \label{eq:ODE-titrationOscillator}
\begin{align}
\dot{x} ={}& \beta^{\rm eff}_\text{X}\l(x\r) - \delta_X x- \alpha x y,\\
\dot{y} ={}& \beta^{\rm eff}_\text{Y}\l(x\r) - \delta_Y y - \alpha x y,
\end{align}
\end{subequations}
where the effective synthesis rate is expressed in terms of the quasi-stationary distribution of the promoter state \citep{angelesHillEquationRevisited1997,linEfficientAnalysisStochastic2017}:
\begin{equation}
    \beta_W^\text{eff}(x) = \beta_W^F + \l(\beta_W^B - \beta_W^F\r) \frac{\l(x\kappa_W /\theta_W \r)^N }{\sum_{i=0}^N \l(x\kappa_W /\theta_W \r)^i}.
\end{equation}
Here, $W\in \l\{X,Y\r\}$. This quasi-stationary simplification is appropriate because the binding and unbinding rates used in our simulations are faster than protein degradation. These sets of ordinary differential equations (ODEs) describe the time-evolution of mean concentrations in the absence of stochastic noise. For each cell with given biophysical parameters and initial conditions, we used standard ODE solvers to simulate the deterministic time-evolution of both the repressilator and titration oscillator. 

\subsection{Bifurcation and stochastic amplification}

Both deterministic dynamical systems Eqs.~\eqref{eq:ODE-repressilator} and ~\eqref{eq:ODE-titrationOscillator} admit perpetually oscillatory solutions (known as limit cycles) for a suitable set of parameters; see Figs.~{\figCircuitTrajectory}C and {\figCircuitTrajectory}E for deterministic limit cycles in the repressilator and titration models. These limit cycles are often stable such that trajectories starting at any point nearby the limit cycle will be ``attracted'' to the limit cycle and the long-time behavior will be arbitrarily close to the flow on the limit cycle. However, most parameter sets in the repressilator and titration models are non-oscillatory and their dynamics will converge to a stable fixed point; see Figs.~{\figCircuitTrajectory}G and {\figCircuitTrajectory}I. As one gradually and continuously changes the parameter(s) of the models, the dynamics can switch abruptly from one qualitative behavior (e.g.~not oscillatory) to another (e.g.~oscillatory). The qualitative change in dynamics at a critical parameter is called a bifurcation \citep{strogatz2014,arnold1999BifurcationTheoryCatastrophe}. The challenge for the evolution of biological oscillators is that most starting parameters are non-oscillatory and that the desired dynamic phenotype is invisible until the population crosses the bifurcation point. 

Stochastic simulation of the repressilator and titration models with the same parameters show two important differences. First, stochastic noise perturbs the limit cycle of oscillatory parameters and affects the period, amplitude, and phase of oscillation; see {\figCircuitTrajectory}D and {\figCircuitTrajectory}F. This will negatively impact the fitness of each oscillator. Second, stochastic noise abolishes the abrupt nature of the bifurcation seen in the deterministic system. For example, stochastic simulation still exhibits oscillatory-like dynamics in the parameter regime where deterministic dynamics are non-oscillatory; see {\figCircuitTrajectory}H and {\figCircuitTrajectory}J. This well-known phenomenon is called \emph{stochastic amplification} \citep{alonso2007StochasticAmplificationEpidemics,bolandLimitCyclesComplex2009}. Below, we will show that stochastic amplification accelerates the evolutionary process of finding oscillatory solutions and crossing bifurcation points. 

\subsection{Evolutionary algorithm} \label{sec:evolution} 

Similar to previous work, we used computer simulation to evolve gene circuits in a haploid, asexual population with non-overlapping generations \citep{gomezschiavon2019}. A schematic diagram of the process can be seen in Fig.~{\figEvolAlgo} along with the corresponding pseudo-code in the Appendix. Briefly, our population of $M$ cells all had the same regulatory network (repressilator or titration oscillator), but each cell had a different set of biophysical parameters $\theta$ whose values represented the underlying ``genotype''.  Each generation, we simulated the deterministic or stochastic dynamics of each cell with its underlying parameters and initial concentrations, and scored its fitness based on whether the dynamic was oscillatory. The next generation of cells were chosen from the top fraction ($\phi$) of the fittest parents and these cells inherited the parental biophysical parameters (i.e., genotypes) with some probability to increase or decrease the biophysical parameter values by some amount (i.e., genetic mutation). In all evolutionary simulations, we only considered population size $M=100$ and strong selection pressure $\phi=0.1$ (only the top 10\% of the population survive to reproduce the next generation). The rest of the evolutionary parameters, both here and below, were variable and their specific values are listed in each figure caption.

\begin{figure}[ht]
\centering
\includegraphics[width=0.9\linewidth]{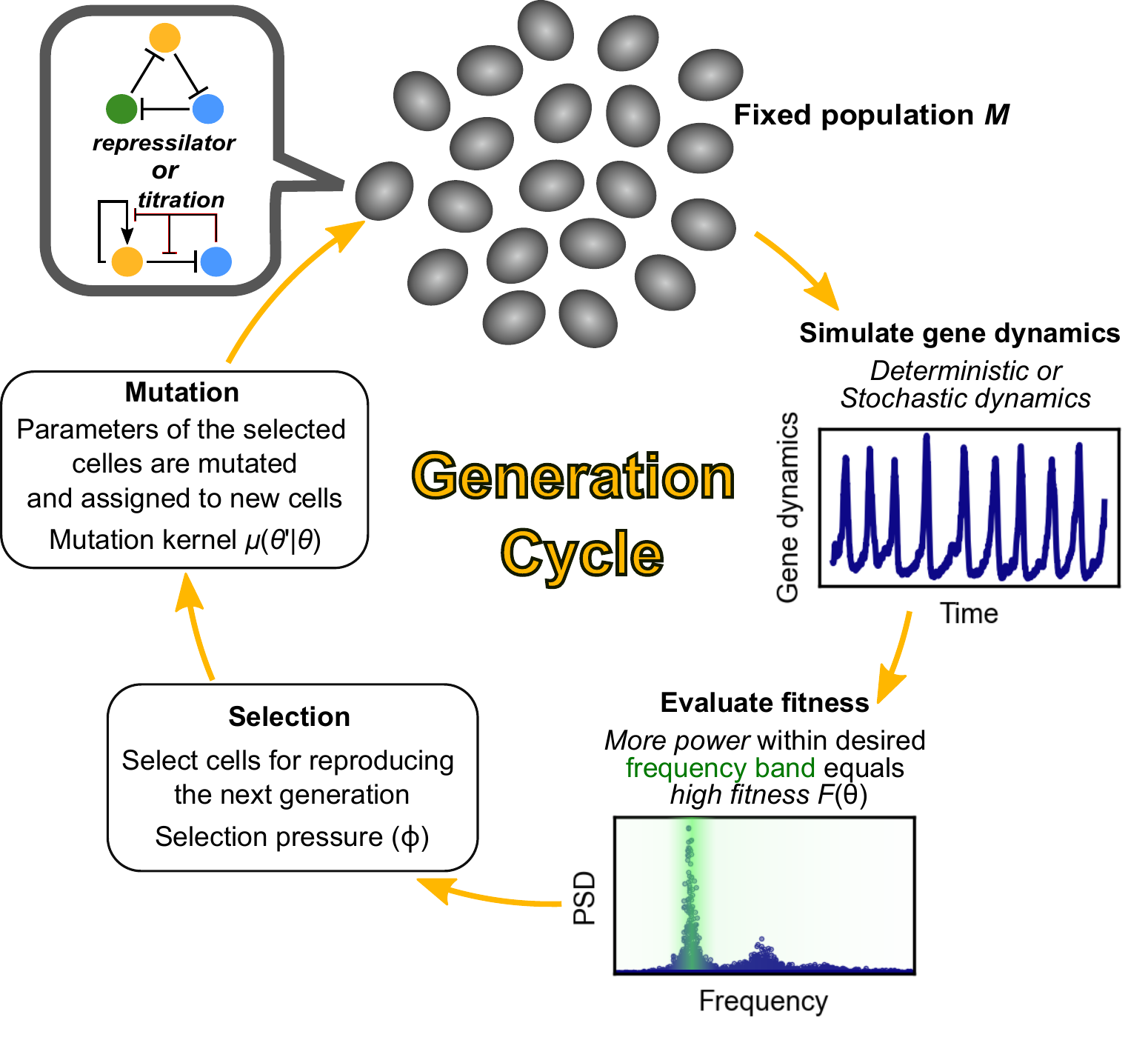}
\caption{{\bf Evolutionary algorithm}. Each generation, we simulated the deterministic or stochastic gene network dynamics (``phenotype'') of the repressilator or titration-based oscillator in each cell given each cell's parameters $\theta$ (``genotype'').  We performed a power spectral density (PSD) analysis of each cell's network dynamics to calculate ``fitness'' ($F$), such that those cells with large amplitude oscillations at a target frequency are fitter cells. We then selected the top fraction ($\phi$) of fittest cells, followed by random replication to replace the remaining population. Each replicated cell inherited the genotype (parameters $\theta$) of the selected parent with a probability of mutation $\mu(\theta^{'} \vert \theta)$ to each parameter.  }
\label{fig:Figure2}
\end{figure}

\subsubsection{Fitness and selection of biological oscillators in a noisy world}

We considered the example of circadian clocks to construct a reasonable fitness function for the evolution of oscillators. The circadian clock plays a role in time-keeping and internalizes the external light-dark intensity. Thus, the highest fitness should occurs when the oscillator frequency $f$ matches that of the external cycle $f_\text{target}$. Furthermore, evolution should select for large amplitude oscillations that are sufficient to match the range of the external signal and drive downstream biochemical processes. 

To this end, we used a Fourier transform of the gene network dynamics to measure the power-spectral density (i.e., frequency and amplitude) of each cell given its set of biophysical parameters. Each generation, we simulated the trajectories of gene dynamics of each cell for total time $T$ and we sampled $\l\{x(t_i)\r\}_i$ (as in Eqs.~\eqref{eq:ODE-repressilator} and \eqref{eq:ODE-titrationOscillator}, and $N_X(t_i)/\Omega$ in the stochastic models) at $10^4$ uniform time intervals over the total time. We then performed a discrete Fourier transform on the sampled time series between $t=T/2$ and $T$ (i.e., $5\times 10^3$ snapshots) so that any transient contribution of initial conditions was negligible. In the deterministic models, the initial concentration was set at $(x,y,z)=(0.1,0,0)$ for the repressilator and $(x,y)=(0,0)$ for the titration oscillator. In the stochastic model, the initial configuration of $(N_X, N_Y)$ was set at the nearest integers to $(0.1\Omega, 0, 0)$ in the repressilator model, and the initial configuration of $(N_X, N_Y, s_X, s_Y)=(0,0,0,0)$ in the titration oscillator. We tested other initial conditions to confirm that our results do not depend on the choice of the initial conditions. The Fourier transform converts the time series $\l\{x\l(t_i\r)\r\}$ to a power-spectral density $\text{PSD}(f)$, where $f$ is the frequency. To avoid artifacts that arise from using discrete Fourier transformation for PSD analysis, we implemented Welch's method (\texttt{scipy.signal.welch}) to estimate a smooth PSD density. From this $\text{PSD}(f)$, we determined the maximum power $\text{PSD}_{\max}:=\max_f \text{PSD}(f)$ and corresponding frequency $f_{\max}:=\text{argmax}_{f} \text{PSD}(f)$ at the maximum power. We rounded off $\text{PSD}_{\max}<10^{-4}$ to eliminate numerical artifacts caused by the finite sampling of the trajectories.   

Our final fitness function satisfied two objectives. First, those circuits whose maximum power frequency $f_{\max}$ matches that of the external cycle $f_\text{target}$ should have higher fitness.  We modeled this with a Lorentzian function, $F\propto 1/\l(1+\l(f_{\max} - f_\text{target}\r)^2\r)$. Second, the amplitude of oscillation should be large enough to match the external signal and drive downstream biochemical processes.  We modeled this ``large enough'' requirement using a simple Michaelis--Menten law, $F \propto \text{PSD}_{\max}/(1+ \text{PSD}_{\max})$. Our final fitness function combined these two objectives via multiplication: 
\begin{equation}
    F\l(\theta\r):= \frac{\text{PSD}_{\max}}{1+\text{PSD}_{\max}} \left( \frac{1}{1 + \l(f_{\max} - f_\text{target}\r)^2} \right),  \label{eq:fitness}
\end{equation}
Thus, those cells with biophysical parameters $\theta$ that place sufficient power at the desired $f_\text{target}$ will have a fitness closer to 1, whereas non-oscillatory cells (or oscillatory cells with insufficient power or at incorrect frequencies) will have a fitness closer to 0. This same fitness function also accounts for the negative influence of noise on proper timing because noise reduces the power of an oscillator at its natural frequency. 

Each generation, the individual cells were ranked by their fitness and the top $\phi$ fraction of the population were selected at random to reproduce and create the next generation of $M$ cells. The biophysical parameters of each new cell $\theta'$ were inherited from their parent with some probability of mutation or change in value. The mutation kernel, $\mu(\theta'\vert \theta)$, quantifies the probability of the new parameter set $\theta'$ given the selected parameter set $\theta$. We chose a normal distribution, such that in a one-dimensional evolutionary process (only evolving a single parameter), $\theta' \sim \mathcal{N}\l(\theta, \sigma^2\r)$ and in a multi-dimensional process (more than one parameters are simultaneously evolving), $\theta' \sim \mathcal{N}\l(\theta, \text{diag}\l(\sigma_1^2, \ldots \r)\r)$ where $\text{diag}\l(\sigma_1^2, \ldots \r)$ is a non-negative-definite co-variance matrix of the multivariate Gaussian distribution. 

\subsection{Stochastic gene dynamics accelerate the evolution of biological oscillation}

Below, we will use our evolutionary simulation to demonstrate that a population of ``noisy'' cells that start with non-oscillatory biophysical parameters (genotype) will evolve by mutation towards an oscillatory phenotype more quickly than a population of cells with deterministic (i.e., noiseless) gene dynamics.  We will show that this acceleration arises because stochastic gene dynamics generate noise-induced oscillations outside of the oscillatory parameter regime and, most importantly, these noise-induced oscillations create a landscape of increasing fitness in the direction of oscillatory parameters. This fitness landscape guides the population via genetic mutation and selection towards robust oscillations. Conversely, cells with deterministic gene dynamics have no oscillation and no fitness gradient outside the oscillatory regime; thus, the population will randomly diffuse on a flat fitness landscape until some cell mutates enough to cross a bifurcation point where deterministic oscillations begin.

\subsubsection{Repressillator evolution}

\begin{figure}[ht]
\centering
\includegraphics[width=0.75\linewidth,angle=270]{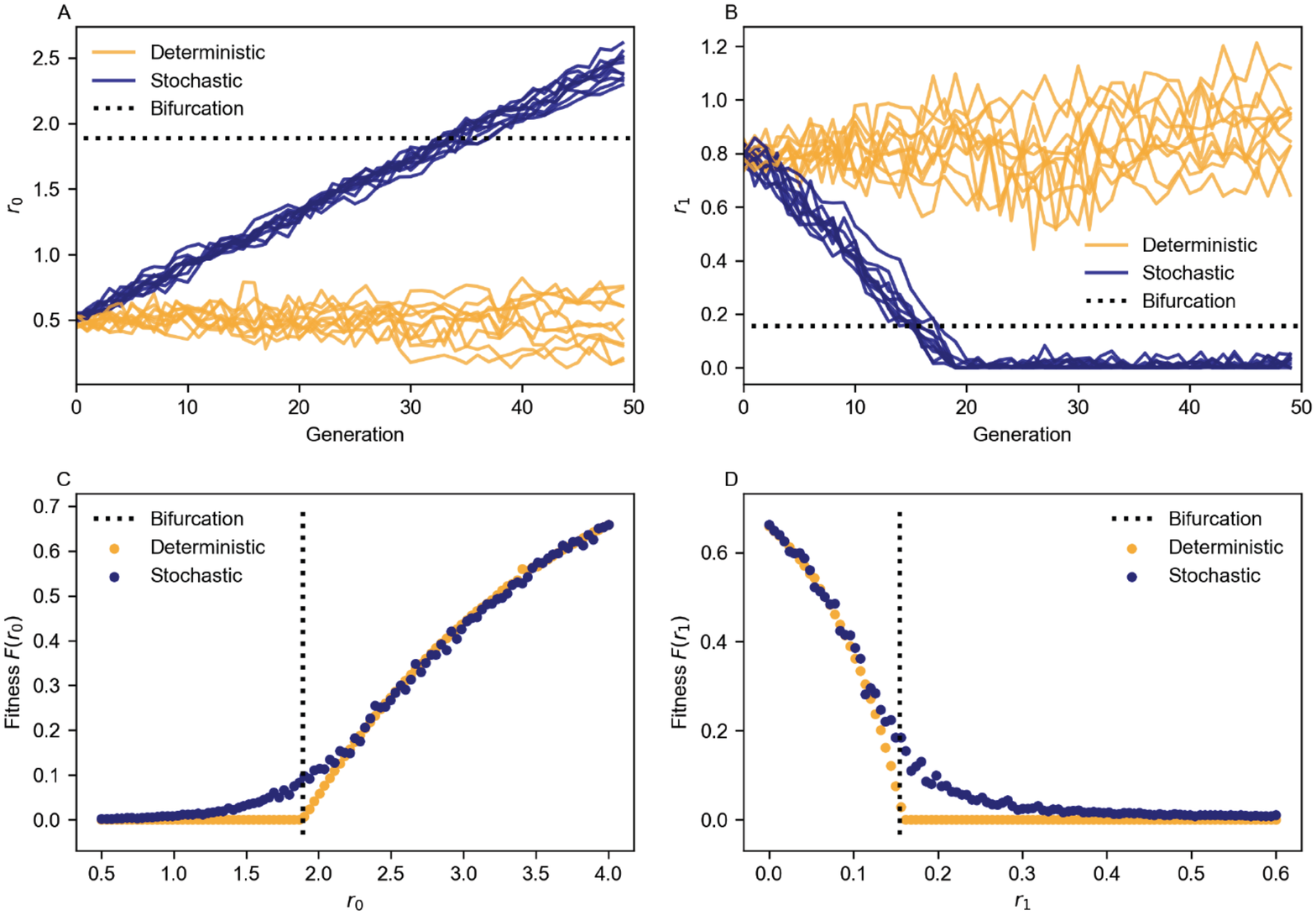}
\caption{{\bf Noise accelerates the evolution of a repressilator by creating a fitness landscape that guides new mutations towards oscillation}. (A,C) Starting from an identical population at $r_0=0.5$ (non-oscillatory phenotype), we ran 10 independent and identical evolutionary simulations for cells with deterministic and stochastic gene dynamics. The horizontal dashed line indicates the value at which a bifurcation occurs and the deterministic system exhibits limit cycles. (A) We plotted $r_0$ of the cell with the best fitness value each generation. (C) We calculated the expected fitness for all values of $r_0$. (B,D) Same as above, but we evolved $r_1$ starting from 0.8 (non-oscillatory phenotype). (B) We plotted $r_1$ of the cell with the best fitness value each generation. (D) We calculated the expected fitness for all values of $r_1$. The evolutionary parameters were $\sigma=0.025$ and $f_\text{target} = 0.3$. We simulated gene dynamics to $T=1000$, with a snapshot every $\Delta t =0.1$. The system size was $\Omega = 500$ for our stochastic simulations.}
\label{fig:Figure3}
\end{figure}

We first performed a one-dimensional evolutionary simulation for the repressilator model shown in Figure 1A. In this set of experiments, we considered $\theta = r_0 \ge 0$ as the mutable parameter and we fixed $r_1=0$. The deterministic dynamics is oscillatory only if $r_0\gtrsim 1.89$ and our population begins the evolutionary simulation in a non-oscillatory parameter regime ($r_0 = 0.5$). We observed a dramatic difference in evolutionary speed and drift between the deterministic and stochastic dynamics (Figure \ref{fig:Figure3}A). To test the generality of this result, we treated $\theta=r_1 \ge 0$ as an adjustable parameter and fixed $r_0=4$. The deterministic dynamics are oscillatory only if $r_1\lesssim 0.16$ and we started the population in a non-oscillatory parameter regime ($r_1 = 0.8$). The evolutionary dynamics, Figure \ref{fig:Figure3}B, are qualitatively similar to that in Fig. \ref{fig:Figure3}A. In both cases, those populations with stochastic gene dynamics robustly identified the direction of the oscillatory parameter regime and consistently improved their fitness, while those with deterministic gene dynamics diffused randomly in the non-oscillatory parameter space. 

These evolutionary results suggest that the fitness landscape sensed by deterministic gene dynamics and stochastic gene dynamics are fundamentally different.  We measured the fitness landscape in each experimental setting in Figs.~\ref{fig:Figure3}C and D, as functions of the adjustable parameter $r_0$ and $r_1$, respectively. The fitness landscape of the deterministic gene dynamics is flat outside the parameter regime admitting oscillatory dynamics, which is consistent with diffusive evolutionary dynamics.  In contrast, the landscape of the stochastic gene dynamics showed a smoothed extension of the fitness landscape into the non-oscillatory parameter regime.  Thus, a population of non-identical genotypes with stochastic gene dynamics will evolve towards oscillatory parameters because the gradient of the fitness landscape will select the fitter variants that are closer to the bifurcation point.  

\begin{figure}[ht]
\centering
\includegraphics[width=0.75\linewidth]{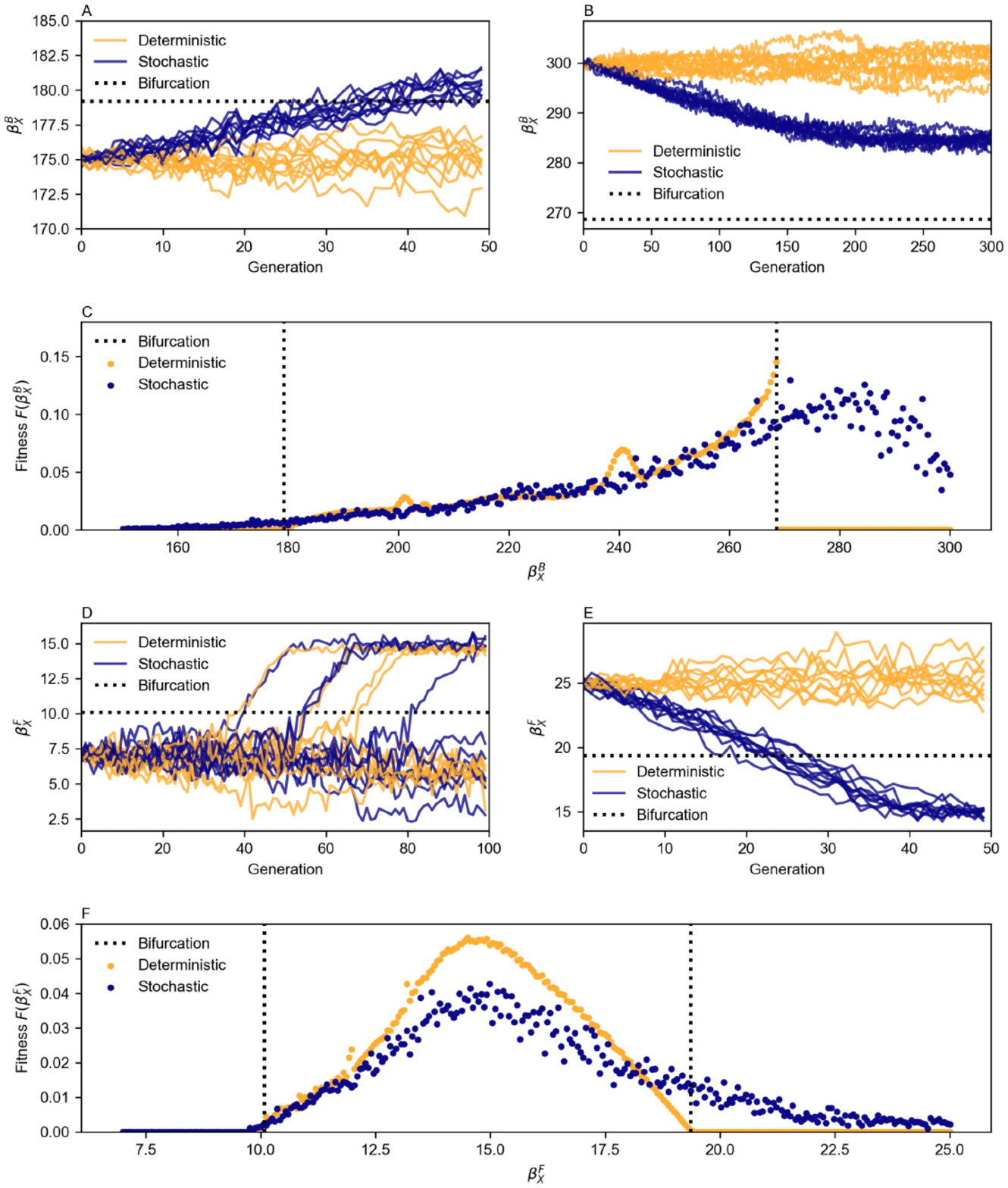}
\caption{{\bf Noise also accelerates the evolution of a titration-based oscillator}. Starting from an identical population at (A) $\beta_X^B=175$ and (B) $\beta_X^B=300$ (non-oscillatory phenotypes), we ran 10 independent and identical evolutionary simulations for cells with deterministic and stochastic gene dynamics. The horizontal dashed line indicates the value at which a bifurcation occurs and where the deterministic system exhibits limit cycles. (A-B) We plotted $\beta_X^B$ of the cell with the best fitness value each generation. (C) We calculated the expected fitness for all values of $\beta_X^B$. (D-F) Same as above, but we evolved populations starting from (D) $\beta_X^F=7.5$ and (E) $\beta_X^F=25$ (non-oscillatory phenotypes). (D-E) We plotted $\beta_X^F$ of the cell with the best fitness value each generation. (F) We calculated the expected fitness for all values of $\beta_X^F$. The evolutionary parameters were $\sigma=0.02$ and $f_\text{target} = 1.5$. We simulated gene dynamics to $T=50$, with a snapshot every $\Delta t =0.005$. The system size was $\Omega = 1000$ in our stochastic simulations.}
\label{fig:Figure4}
\end{figure}

\subsubsection{Titration oscillator evolution}

To verify that our results are not particular to the repressilator, we also evolved a population of titration oscillators because the positive and negative feedback loops generate more complex dynamics, and these could affect stochastic amplification.  We first ran a one-dimensional evolutionary algorithm for two different parameter cases: (1) an evolvable $\beta_X^B \ge 0$ with a fixed $\beta_X^F=12$, and (2) an evolvable $\beta_X^F \ge 0$ with a fixed $\beta_X^B=200$. In the first case, there are two bifurcation points and the deterministic dynamics are oscillatory only if $179.20 \lesssim \beta_X^B \lesssim 268.62$. In the second case, there are two bifurcation points and the deterministic dynamics are oscillatory only if $10.08 \lesssim \beta_X^F \lesssim 19.36$. In both cases, we started the evolutionary processes in distinct, non-oscillatory regimes outside the low and high bifurcation points.

\begin{figure}[ht]
\centering
\includegraphics[width=0.9\linewidth]{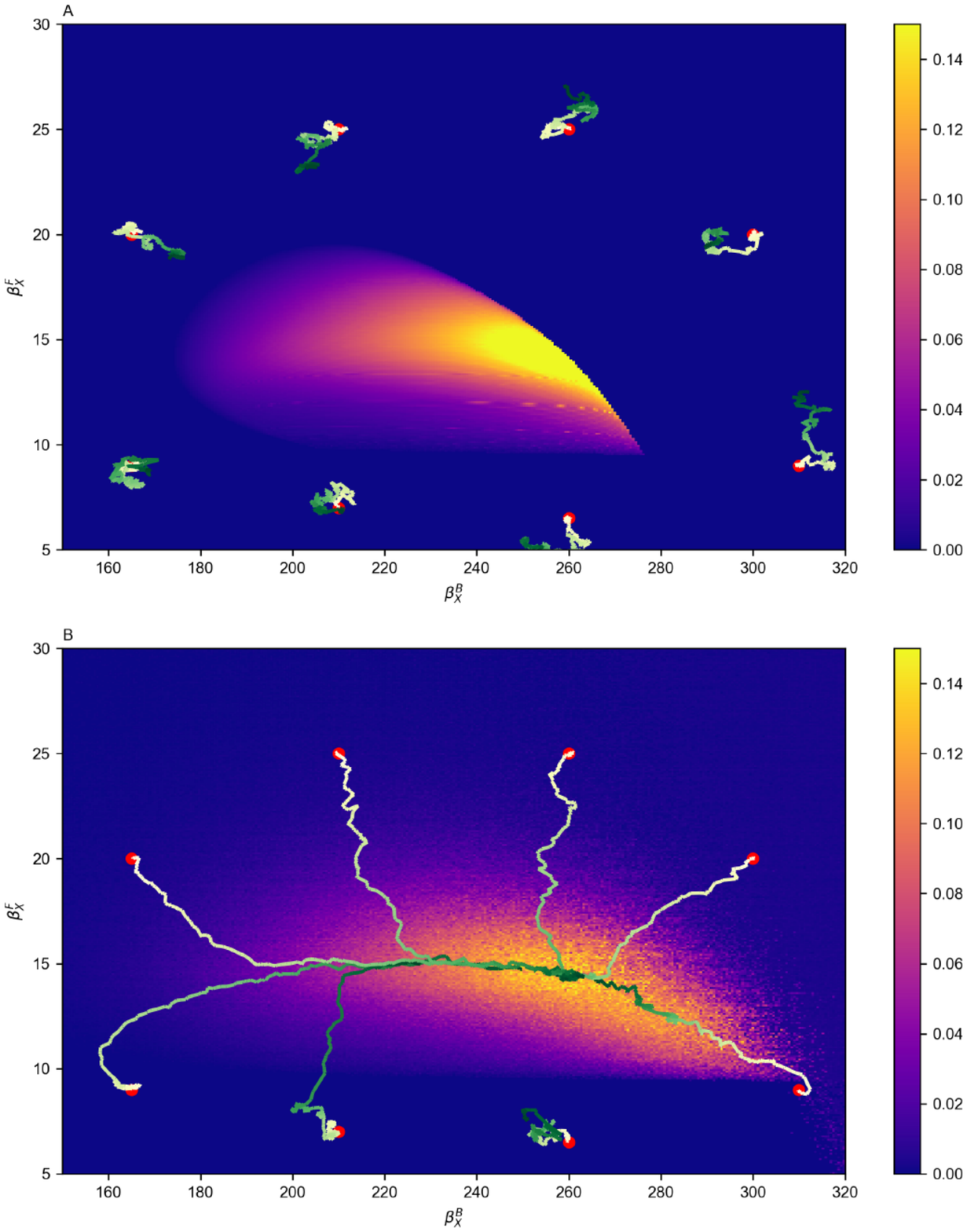}
\caption{{\bf The fitness landscape of a titration-based oscillator is extended and smoothened by stochastic amplification}. (A) Fitness landscape and evolutionary trajectories of populations with (A) Deterministic gene dynamics, and (B) Stochastic gene dynamics. The evolutionary parameters were identical to Figure 4, except that $\text{diag}\l(0.01, 0.4\r)$, which accounts for distinct timescales of evolution for $\beta_X^F$ and $\beta_X^B$ (see Fig.~\ref{fig:Figure5}C and F). We simulated 200 generations.}
\label{fig:Figure5}
\end{figure}

We again observed that stochastic gene dynamics significantly accelerate the evolutionary processes of finding an oscillatory parameter regime, with the exception of the region below $\beta_X^F \approx 10$ (Fig.~\ref{fig:Figure4}D). In agreement with our previous results, the fitness landscapes (Figs.~\ref{fig:Figure4}C and F) extend into regions adjacent to the bifurcation points due to stochastic amplification.  This is true even for the region below $\beta_X^F \approx 10$, where there is still a small extension of the fitness landscape near the bifurcation point (Fig.~\ref{fig:Figure4}F). To verify that our results were not a consequence of evolving one parameter at a time, we simulated a two-dimensional evolutionary processes of the titration model, where $\theta=\l(\beta_X^F, \beta_X^B\r)$ are both mutable. We initialized our population at eight different parameter settings outside the parameter regime permitting deterministic oscillatory dynamics. Seven out of eight populations with stochastic gene dynamics rapidly evolved into the oscillatory parameter regime (Fig.~\ref{fig:Figure5}A), whereas the deterministic populations were still randomly diffusing in the non-oscillatory parameter space.  As expected, the fitness landscape of the stochastic populations (Fig.~\ref{fig:Figure5}B) was broader and provided a smooth gradient towards oscillatory parameters, when compared to the flat fitness landscape of deterministic populations in Fig.~\ref{fig:Figure5}A. Thus, smooth extensions of the fitness landscape induced by the stochastic gene dynamics is a robust result. Some directions showed a larger fitness landscape extension when compared to other directions, suggesting that some parameters are more susceptible to the effects of noise-induced oscillation and accelerated evolution.  

Within the oscillatory parameter regime, the fitness of the stochastic gene dynamics is lower than that of deterministic dynamics (compare deterministic and stochastic fitness landscapes in Fig.~\ref{fig:Figure4}C and F). The reduced fitness of stochastic dynamics is a consequence of the destructive properties of noise on precise time-keeping or frequency matching. As such, there is an interesting trade-off between stochastic and deterministic dynamics: outside the oscillatory parameter regime, the former is better in ``sensing'' the direction towards oscillatory parameters, but once inside the oscillatory parameter regime, the latter is better in generating coherent oscillation. 

\subsubsection{Noise as an evolvable parameter during the evolution of oscillation}

So far, we ran evolutionary simulations in which \emph{all the cell} circuits are either modeled using deterministic or stochastic gene dynamics. However, Figure \ref{fig:Figure4} showed that the deterministic gene dynamics can be more advantageous in the oscillatory parameter regime because they are more coherent. It is natural to ask the following questions: What is the evolutionary process if each cell is allowed to change its noise strength?  Would the population first select stochastic gene dynamics for better fitness and then more efficiently evolve towards oscillatory parameter space, with a later switch to deterministic dynamics for a more coherent oscillation?

To test this hypothesis, we considered a three dimensional evolutionary process, where the evolvable parameters are $\theta = \l(\beta_X^F, \beta_X^B, \log_{10} \Omega \r)$. Recall that $\Omega$ is the system-size parameter controlling the strength of the intrinsic noise in the stochastic gene dynamics; larger $\Omega$ corresponds to smaller intrinsic noise.  We allowed each cell to have its own $\Omega$ to be mutated and selected by the evolutionary processes. The cells could evolve their volume in the range of $10^3 < \Omega < 10^5$. At $\Omega \ge 10^5$, we switched to a deterministic simulator (Det) because the stochastic simulator became inefficient and time-consuming. 

\begin{figure}[ht]
\centering
\includegraphics[width=0.85\linewidth]{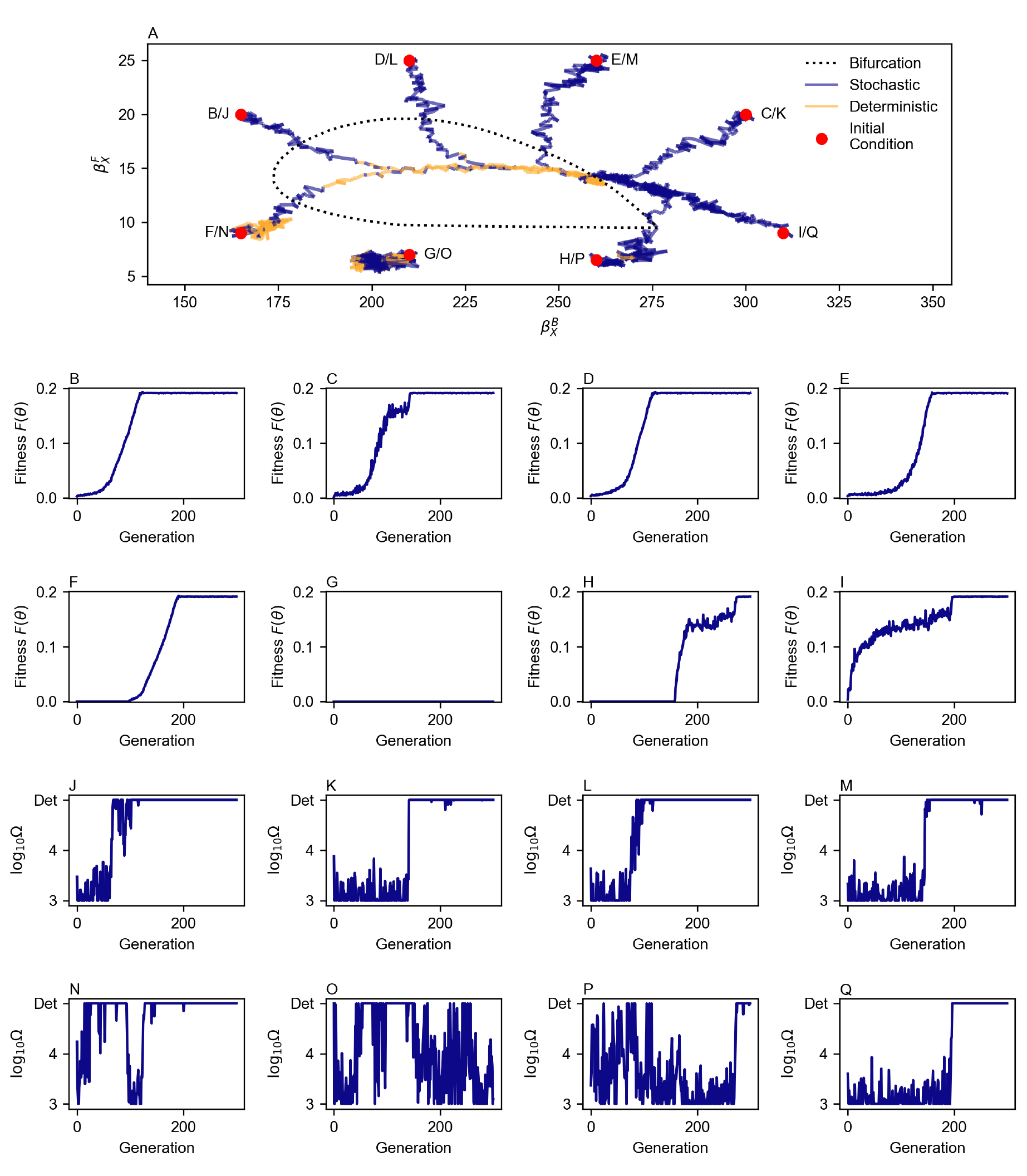}
\caption{{\bf The evolutionary process selects for noise-induced oscillation outside the bifurcation boundary}. (A) Evolutionary trajectories of all populations. We plot the parameters of the fittest member of each evolving population, including system size ($10^3 < \Omega < 10^5$ is purple, whereas $\Omega \ge 10^5$ is yellow and deterministic). The alphabetic labels are associated with the plots of (B-I) fitness $F(\theta)$ and (J-Q) system size $\Omega$ over evolutionary time. The evolutionary parameters were identical to Figure 5, except that $\text{diag}\l(0.01, 0.4, 0.1\r)$ and all populations started with a system size of $\Omega=10^4$.  Note that the mutation of the system size $\Omega$ takes place in the $\log_{10}$ space, leading to a log-normal mutation kernel of $\Omega$. We simulated 300 generations.}
\label{fig:Figure6}
\end{figure}

Figure \ref{fig:Figure6}A shows a two-dimensional projection of the evolutionary trajectories. For most trajectories, there were three evolutionary phases. In the first phase, the cells immediately evolved to a regime with lower $\Omega$ and increased stochastic noise. These noisier cells exhibited better noise-induced oscillations and increased their fitness in the non-oscillatory parameter regime. Importantly, these noisier cells could sense a smooth, increasing fitness gradient. Thus, in the second phase, cells responded to this fitness gradient and evolved towards the oscillatory parameter regime. This effect can be seen in the corresponding plots of fitness (Figure \ref{fig:Figure6}B-6I) and system size (Figure \ref{fig:Figure6}J-6Q) over generation time, where cells kept noise high and evolved up the fitness landscape. Once cells crossed the bifurcation point into the oscillatory regime, noise became more detrimental to overall fitness. Thus, cells enter a third and final phase where they evolved to increase system size to produce more coherent oscillation via deterministic dynamics. Most populations followed these three evolutionary phases; however, there were exceptions for the region near or below $\beta_X^F \approx 10$, where the stochastic dynamics landscape was flatter (Figure \ref{fig:Figure6}F-6H) and the populations tended to diffuse randomly in all three parameters, including $\Omega$. However, we note that two of those populations eventually diffused to regions of parameter space where they could sense the noise-induced fitness gradient, at which point they entered the second phase (more noise, generate and follow the fitness gradient) and third phase (less noise, more coherent oscillation, increased fitness). 

\subsection{Mechanisms of noise-induced oscillation and stochastic amplification} \label{sec:noise-induced}

\begin{figure}[ht]
\centering
\includegraphics[width=0.8\linewidth]{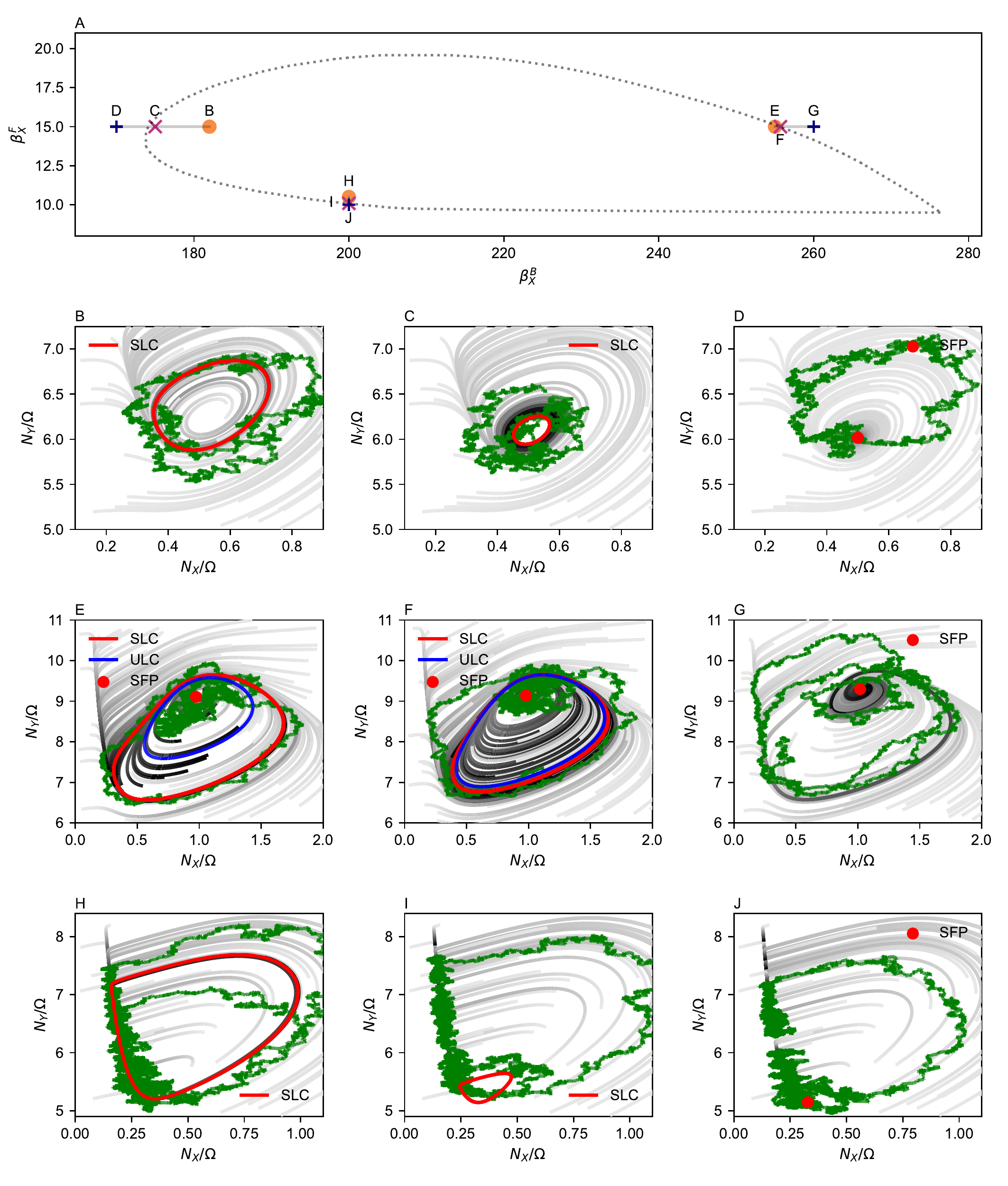}
\caption{{\bf Mechanisms of noise-induced oscillation}. (A) We considered three parameter sets that cross different bifurcation boundaries. Each parameter set contains one point inside the boundary, one inside-but-close to the boundary, and one outside the boundary. The system size was set to $\Omega=1000$ for stochastic simulation. (B-D) We plot the phase portraits of the system along with the stochastic and deterministic trajectories for fixed $\beta_X^F=15$ and variable $\beta_X^B=182$, $175$, and $170$ respectively. (E-G) Phase portraits and dynamics of the system for fixed $\beta_X^F=15$ and variable $\beta_X^B=255$, $255.7$, and $260$ respectively. (H-J) Phase portraits and dynamics of the system for fixed $\beta_X^B=200$ and variable $\beta_X^F=10.5$, $10.09$, and $10$ respectively. SLC = stable limit cycle, ULC = unstable limit cycle, SFP = stable fixed point.}
\label{fig:Figure7}
\end{figure}

To provide a mechanistic understanding of why stochastic gene expression accelerates the evolutionary processes, we plotted a phase portrait and stability analysis of the deterministic system for three parameter sets that cross different types of bifurcation boundaries, and we super-imposed the corresponding stochastic dynamics trajectories; see Figure \ref{fig:Figure7}.

\subsubsection{Simple Hopf bifurcation and noise-induced oscillations}
The first parameter set shown in Figure ~\ref{fig:Figure7}B-D traverses a simple Hopf bifurcation \citep{strogatz2014}. For parameters inside the bifurcation boundary, the deterministic trajectory converges to a globally stable limit cycle (SLC); see Fig.~\ref{fig:Figure7}B. As the critical parameter moves towards the boundary, the amplitude of the deterministic limit cycle continuously shrinks to zero (Fig.~\ref{fig:Figure7}C). Once beyond the boundary, the system no longer has limit cycles (Fig.~\ref{fig:Figure7}D) and all trajectories spiral to a stable fixed point (SFP).  As expected, stochastic gene dynamics perturb deterministic limit cycles and have less coherent oscillation inside the boundary, which reduces fitness (see green trajectories in Figs.~\ref{fig:Figure7}B-C). However, outside the bifurcation boundary, stochastic gene dynamics has a positive influence on fitness by perpetually ``kicking'' the state out of the neighborhood of SFP and relaxing back to the stable fixed point with a spiral trajectory (i.e. under-damped oscillation). The two forces, the former ``fluctuation'' and the latter ``relaxation'', are responsible for generating noise-induced oscillations; see stochastic time series (Fig.~\ref{fig:FigureS1}G) and frequency mode in the PSD (Fig.~\ref{fig:FigureS2}G). This well-known phenomenon is called a stochastic Hopf bifurcation \citep{hoppe1996StochasticHopfBifurcation,ma2012StochasticHopfBifurcation,arnold1999StochasticBrusselatorParametric,li2019StochasticBifurcationAnalysis,simpson2021StochasticHopfBifurcations}. The noise-induced oscillations have a frequency whose value is related to the imaginary part of the eigenvalues of the Jacobian at the fixed point underlying deterministic dynamics. We note that the noise-induced oscillations of the repressilator have the same origin and mechanism as the parameter set described here because the repressilator has a simple Hopf bifurcation.

\subsubsection{Bautin bifurcation and noise-induced oscillations}
The second parameter set shown in Figure ~\ref{fig:Figure7}E-G shows the signature of a generalized Hopf bifurcation, also known as a Bautin bifurcation \citep{guckenheimer2002NonlinearOscillationsDynamical}. Near the bifurcation and inside the oscillatory parameter regime, Figs.~\ref{fig:Figure7}E,F there exist an unstable fixed point, an unstable limit cycle (ULC), and a stable limit cycle (SLC). The SLC and ULC partition the entire phase space, such that any initial state inside or outside the ULC will eventually converge to the SFP or SLC, respectively. As the critical parameter moves towards the bifurcation point, the ULC grows larger (Fig.~\ref{fig:Figure7}F) and eventually collides with and annihilates the SLC at the bifurcation point.  The amplitude of the oscillation will suddenly vanish as the finite-size SLC is destroyed by the ULC. This is in contrast to the simple Hopf bifurcation, whose amplitude continuously converged to zero at the bifurcation. Such a sudden annihilation is reflected in Fig.~\ref{fig:Figure4}C, where the deterministic fitness function has a discontinuity at the Bautin bifurcation.

\begin{figure}[ht]
\centering
\includegraphics[width=0.8\linewidth]{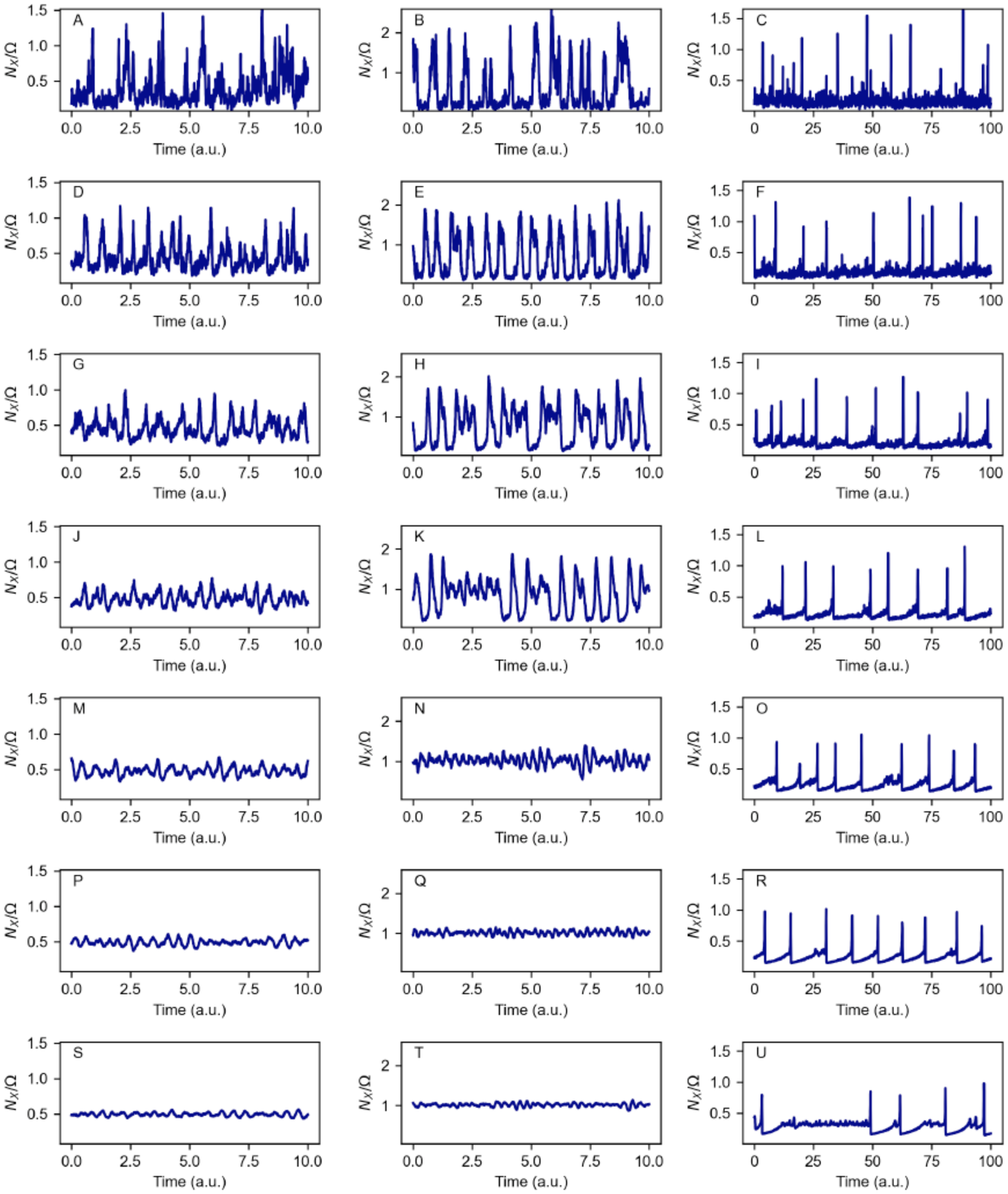}
\caption{{\bf Stochastic dynamics of the titration-based oscillator as a function of intrinsic noise.} The parameter sets are $(\beta_X^F,\beta_X^B)=(15,170)$ (left column), $(15,260)$ (middle column), $(9.8, 200)$ (right column), and correspond to Panels D, G, J in Figure 7, respectively. We adjusted the strength of intrinsic noise by changing the system size parameter $\Omega$. From the top to the bottom row, we set $\Omega=10^2$, $10^{2.5}$, $10^{3}$, $10^{3.5}$, $10^{4}$, $10^{4.5}$, and $10^{5}$. A system with larger $\Omega$ has smaller intrinsic noise. The dynamics with these parameter sets exhibit \emph{noise-induced oscillations}: the stochastic dynamics exhibit oscillatory behavior despite the parameters sets do not admit deterministic oscillations. }
\label{fig:FigureS1}
\end{figure}

Beyond the bifurcation point, the flow in the phase portrait still retains ``pseudo SLC'' dynamic, a dominant one-dimensional convergence path that resembles the annihilated SLC (Fig.~\ref{fig:Figure7}G). As the system spirals to its SFP with a spiral trajectory (i.e. under-damped oscillation), noise will occasionally kick it to the pseudo SLC basin and generate large amplitude noise-induced oscillations; see stochastic trajectory in Fig.~\ref{fig:FigureS1}E. The pseudo SLC basin and associated large amplitude noise-induced oscillations are more clearly visible at slightly larger system size (Fig.~\ref{fig:FigureS1}K), where the intrinsic noise is smaller and cannot reliably kick the system to pseudo SLC every time the state is near the SFP. In this regime, the state stays near the SFP for a random time before noise induces a larger cycle along the pseudo SLC.  If noise becomes too small (larger $\Omega$), then the system is unable to access the pseudo SLC (Fig.~\ref{fig:FigureS1}N, Q, T). Conversely, if noise is too large (smaller $\Omega$), then noise-induced oscillations become less coherent and more variable (Fig.~\ref{fig:FigureS1}B). This suggests that a moderate amount of noise generates the largest peak in PSD (higher amplitude, more coherent noise-induced oscillations); see Fig.~\ref{fig:FigureS2}E,H.  We return to this phenomenon, generally known as ``coherence resonance'', in the excitable case study below. 

\subsubsection{Excitation-relaxation and coherence resonance}
\begin{figure}[ht]
\centering
\includegraphics[width=0.75\linewidth]{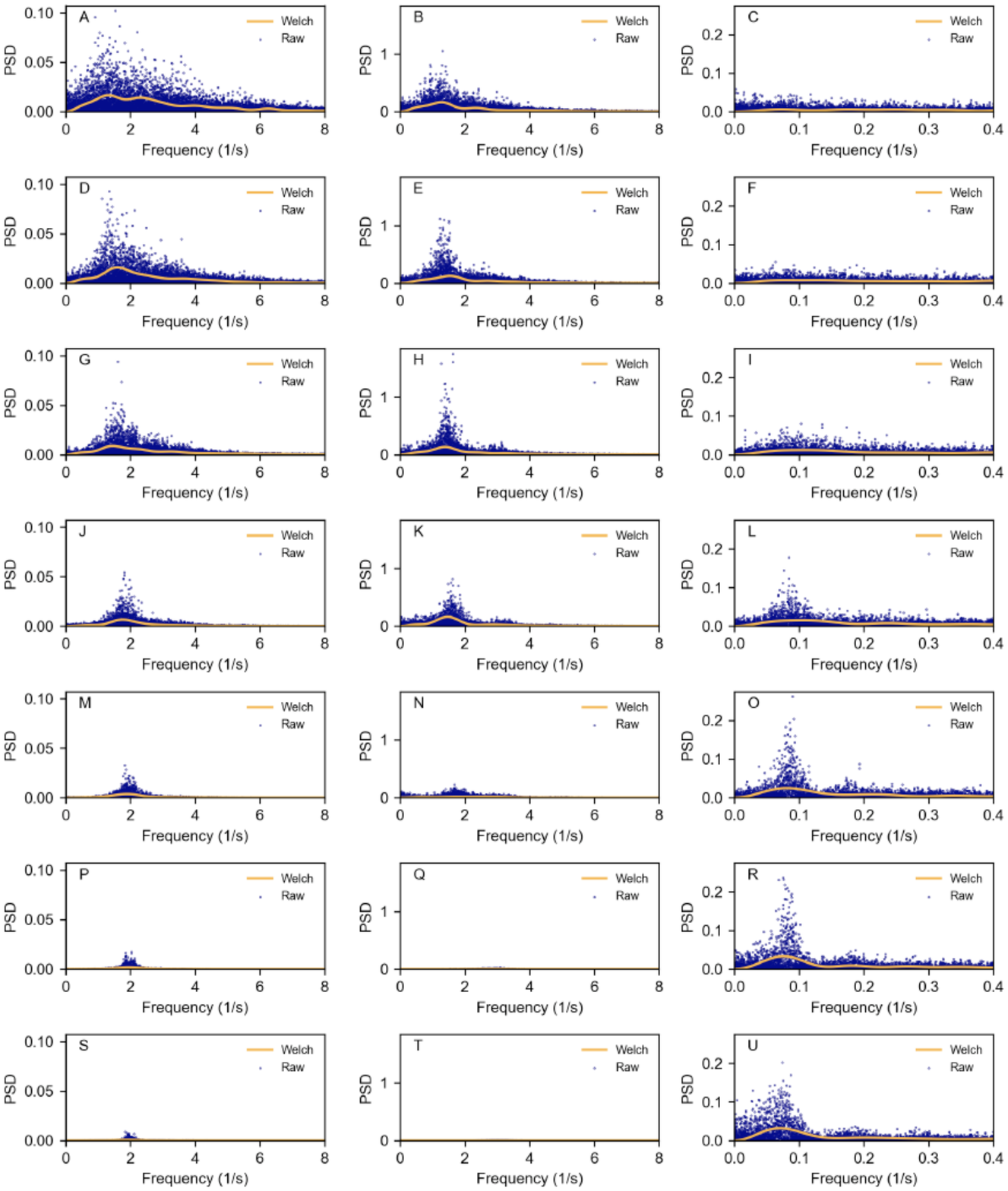}
\caption{{\bf Power-spectral analysis of the titration-based oscillator as a function of intrinsic noise.} The parameter sets are $(\beta_X^F,\beta_X^B)=(15,170)$ (left column), $(15,260)$ (middle column), $(9.8, 200)$ (right column), and correspond to Panels D, G, J in Figure 7, respectively. From the top to the bottom row, we set $\Omega=10^2$, $10^{2.5}$, $10^{3}$, $10^{3.5}$, $10^{4}$, $10^{4.5}$, and $10^{5}$.  We collected $2^{18}$ snapshots for performing power-spectral density analysis after discarding the first $10^4$ transient, initial-condition-dependent snapshots. For the first two parameter sets, we chose $\Delta t=0.005$. For the last parameter set, we chose $\Delta t=0.05$ because the noise-induced oscillations have a longer timescale. We also plot the smoothed density estimation obtained by Welch's method using $5\times 10^3$ snapshots after discarding $5\times 10^3$ initial snapshots, as we did for the evolutionary processes. This shows that the density estimation of the shorter time series by Welch's method reasonably captures the behavior of the raw PSD using a longer time series. Segments of the time series are visualized in Supplementary Fig.~\ref{fig:FigureS1}.}
\label{fig:FigureS2}
\end{figure}

Similar to the first parameter set, the deterministic flow analysis of the third parameter set shows a Hopf bifurcation where the amplitude of the SLC continuously decreases to zero as the parameters approach the bifurcation boundary (Figs. \ref{fig:Figure7}H and I). However, unlike the first parameter set, the SFP is non-oscillatory and over-damped for parameters beyond the boundary. The phase portrait exhibits a flow where trajectories can have long, excitable excursions that relax back to the globally-stable SFP; see Fig.~\ref{fig:Figure7}J. A consequence of this flow is that a stochastic ``kick'' in the right direction will induce a large amplitude excursion in phase space. If noise is too large (small $\Omega$), then it generates incoherent excursions; see stochastic trajectories in Fig.~\ref{fig:FigureS1}C and F. If noise is too small (large $\Omega$), then the system is incapable of escaping the SFP and being regularly kicked into an excitable excursion; see Fig.~\ref{fig:FigureS1}U. This noise-induced phenomenon for excitation-relaxation systems is known as ``coherence resonance'' \citep{pikovsky1997CoherenceResonanceNoiseDriven}, where an optimal level of noise induces coherent noise-induced oscillations. This is best seen in the PSD of the excitation-relaxation system as a function of intrinsic noise (system size), where there is a well-defined mode at $\sim 0.09$ that peaks at intermediate levels of noise (Fig.~\ref{fig:FigureS2}O). We note that the second parameter set also exhibits a well-defined mode that appears to peak at intermediate levels of noise (Fig.~\ref{fig:FigureS2}H), suggesting that parameters outside the Bautin bifurcation exhibit coherence resonance. 
    
\section{Discussion} \label{sec:discussion}

The goal of this study was to understand the impact of biochemical noise, an intrinsic property of living systems, on the evolution of a dynamical phenotype (i.e., oscillation) in gene regulatory networks. Ideally, this could be addressed using laboratory evolution of gene circuits in bacteria or yeast. However, such an approach is infeasible because (1) the biophysical parameters of gene circuits are mostly unknown and unmapped to the genome sequence, (2) the timescale of laboratory evolution is too long, and (3) the level of biochemical noise in a living system is not easily tunable. Here, we used computer simulation to mimic the process of evolution of simple gene circuits, where biophysical parameters are randomly mutated and fitter variants that oscillate better and more robustly are selected each generation. Importantly, this \emph{in silico} approach allowed us to directly measure the effects of gene expression noise on the process of evolution because we can vary the level of intrinsic noise by changing the system size in the Gillespie simulation of stochastic gene expression dynamics. 

We evolved two different gene circuit architectures, the repressilator and a titration-based oscillator. All circuits started with initial biophysical parameters (genotype) where the underlying deterministic dynamic is non-oscillatory and converges to a stable fixed point (phenotype). Gene expression noise accelerated the speed at which evolution finds parameters that oscillate robustly, both for the repressilator (Fig. 3) and titration-based oscillator (Figs. 4, 5). This counter-intuitive effect arises because stochastic noise perturbs the circuit into a oscillatory-like dynamics with non-zero fitness.  Importantly, these noise-induced oscillations generate a fitness landscape whose increasing gradient guides evolution across the bifurcation into a region of parameter space with robust oscillations. We tested the idea that noise is adaptive by allowing cells to evolve their level of gene expression noise. All circuits started with deterministic dynamics (no noise) at the beginning of the evolutionary simulation. Most evolving cells immediately increased their gene expression noise and generated noise-induced oscillations with non-zero fitness. This was followed by the accelerated evolution of biophysical parameters towards the bifurcation point via the fitness gradient (Fig. 6). Once across the bifurcation point, cells evolved back to deterministic limit cycles because noise reduced the maximum possible fitness. These results show that gene expression noise can be adaptive and, thus, imparts a benefit to living organisms. 

An analysis of the dynamics for non-oscillatory parameters near the bifurcation boundary suggests that the mechanism of noise-induced oscillation is very generic. Remnants of limit cycle-like dynamics (``ghosts'') are still present in the deterministic flow field, and the overall effect of stochastic noise is to continuously perturbs the system into limit cycle-like excursions (Fig. 7). If the noise is too small, then the deterministic flow field dominates and the pseudo-oscillation have small amplitude (see power spectral density, PSD, in Figures S2). If the noise is too large, the limit cycle-like dynamics are strongly perturbed and the PSD frequency peak broadens and flattens. There is an optimal level of noise where pseudo-oscillations have a maximum PSD at some characteristic frequency. This phenomenon is known as ``coherence resonance'' and was first described for excitable systems \citep{pikovsky1997CoherenceResonanceNoiseDriven}. Our results suggest that coherence resonance (i.e., noise-induced oscillations that arise outside the bifurcation boundary at intermediate levels of noise) occurs for other types of bifurcations. Most importantly, coherence resonance generates a non-zero fitness landscape whose gradient guides and accelerates evolution across the bifurcation boundary towards robust, self-sustaining oscillations.

Although our work specifically focused on the evolution of oscillation in gene regulatory networks, we suspect that noise will have similar effects on the evolution of other dynamics (e.g., multi-stability) near bifurcation boundaries. Namely, noise will perturb the dynamical system and reveal remnants of multi-stable-like dynamics, such that it will guide and accelerate the evolution of multi-stability. Evolutionary algorithms are part of a class of algorithms used in Machine Learning and the optimization of information processing networks (e.g. neural networks, electronic circuits). Similar to our results, previous work showed that adding noise to Machine Learning algorithms can significantly improve learning. Existing examples include the probabilistic computational unit built in fuzzy logic \citep{novak2012mathematical}, stochastic inputs in Mixup training \citep{zhang2018mixup,NEURIPS2019_36ad8b5f}, stochastic computational processes in variational inference algorithms \citep{kingmaAutoencodingVariationalBayes2014,salimansMarkovChainMonte2015}, and noisy recurrent neural networks \citep{lim2021noisy}. The common idea in these approaches is that adding noise or ``misinformation'' to the learning processes forces the algorithm to learn more robust solutions. Our work suggests that adding noise to these networks could also accelerate and guide the learning process by revealing ``cryptic'' phenotypes outside the bifurcation boundary. Thus, noise may provide two separate benefits (accelerated learning and robust solutions) for Machine Learning algorithms in general. 

\section{Limitations of the study} \label{sec:limitation}

There are several observations regarding noise-induced oscillations and the evolution of biological oscillators. First, although the fitness landscape extends beyond the deterministic bifurcation boundary, the fitness extension was larger for some parameters compared to others. Cells in those extended regions evolved faster and in more directed fashion compared to populations in less extended regions (Figure 5). This suggests that some regimes of biophysical parameters are more evolvable than others, and that this depends on the underlying circuit topology and associated bifurcations. Thus, not all circuits and populations are equally evolvable and this may constrain the types of oscillators that evolve first and fastest. Second, the efficacy of gene expression noise to accelerate and guide the evolution of oscillators depends on the level of intrinsic noise, the underlying process of evolution, and the distance of the initial population from a bifurcation boundary. For example, non-oscillatory parameters further away from bifurcation will not have a strong gradient and, thus, noise is expected to have less effect on the speed of evolution. Conversely, if the mutation step size is large enough for the parameters to randomly diffuse into robust oscillatory parameter space within a few steps, then noise will have less effect on the speed of evolution. The extent to which noise accelerates and guides evolution of oscillation in natural biological systems remains to be determined. 

\section{Resource availability} \label{sec:resource}
\subsection{Lead contact}
Nicolas E. Buchler, nebuchle@ncsu.edu

\subsection{Materials Availability}
This study did not generate reagents or perform physical experiments. 

\subsection{Data and Code Availability}

The complete source codes of this project are deposited in \href{https://github.com/lanl/In-Silico-Evolution-of-oscillatory-gene-dynamics}{a GitHub repository}. The source codes of this project has been reviewed approved for an open-source release, with a LANL C-number C21109, by the Richard P. Feynman Center for Innovation (FCI) at the Los Alamos National Laboratory.


\section{Acknowledgement} \label{sec:acknowledgement}
This work was supported by R01GM127614 from the National Institute of Health (NEB) and the Laboratory Directed Research and Development Program (LDRD project XX9C) at Los Alamos National Laboratory (YTL). 

\section{Author contributions} \label{sec:authorContributions}
Y.T.L and N.E.B.~conceived the idea of the study, designed \emph{in-silico} experiments, and wrote the manuscript. Y.T.L.~developed the code and performed numerical experiments. 

\section{Declaration of interests} \label{sec:CoI}
The authors declare no competing interests.

\bibliographystyle{elsarticle-harv} 
\bibliography{refs}

\section{Appendix}

\subsection{Stochastic formulation of the repressilator and circadian clock}

The stochastic process can be fully specified by a Chemical Master Equation (CME) describing the evolution of the system's time-dependtnt joint probability distribution \citep{vanKampen,gardiner}. Following the procedures described in \citet{vanKampen,linEfficientAnalysisStochastic2017,lin2019ScalingMethodsAccelerating}, we construct continuous-time Markov chains describing the stochastic reaction events in a finite system ($\Omega < \infty$). For the repressilator model, the joint probability distribution is 

\begin{equation}
    p_{i,j,k}(t):=\pr{N_X(t)=i, N_Y(t)=j, N_Z(t)=k},
\end{equation}
where $N_X(t)$, $N_Y(t)$, $N_Z(t)$ are random variables describing the discrete, non-negative, and integer-valued populations of the species $X$, $Y$, and $Z$ at time $t$, respectively. The CME is a countably infinite set of coupled ordinary differential equations:
\begin{align}
    \dot{p}_{i,j,k} ={}&\Omega \cdot H\l(\frac{j}{\Omega}\r) \cdot \l[ p_{i-1,j,k} - p_{i,j,k} \r] \nonumber \\
    +{}&\Omega \cdot H\l(\frac{k}{\Omega}\r) \cdot \l[ p_{i,j-1,k} - p_{i,j,k} \r] \nonumber \\
    +{}&\Omega \cdot H\l(\frac{i}{\Omega}\r) \cdot \l[ p_{i,j,k-1} - p_{i,j,k} \r] \nonumber \\
    +{}&\delta \cdot \l[ \l(i+1\r) p_{i+1,j,k} - ip_{i,j,k} \r]\nonumber\\
    +{}&\delta \cdot \l[ \l(j+1\r) p_{i,j+1,k} - jp_{i,j,k} \r]\nonumber\\
    +{}&\delta \cdot \l[ \l(k+1\r) p_{i,j,k+1} - kp_{i,j,k} \r] \label{eq:CME-repressilator}
\end{align}
with boundary conditions $p_{i,j,k}(t)=0$ if $i<0$, $j<0$, or $k<0$. As for the titration model, the joint probability distribution is 
\begin{equation}
    p_{i,j,k,l}(t):=\pr{N_X(t)=i, N_Y(t)=j, S_X(t)=k, S_Y(t)=l},
\end{equation}
and the CME is \citep{linEfficientAnalysisStochastic2017}:
\begin{align} 
    \dot{p}_{i,j,k,l} ={}& \l(p_{i-1,j,k,l}- p_{i,j,k,l}\r) \mathbf{1}_{\l\{k<N\r\}} \Omega \beta_\text{X}^F \nonumber \\
{}& +  \l( p_{i-1,j,k,l}- p_{i,j,k,l}\r) \mathbf{1}_{\l\{k=N\r\}} \Omega \beta_\text{X}^B \nonumber \\
{}& +\l(p_{i,j-1,k,l}- p_{i,j,k,l}\r) \mathbf{1}_{\l\{l<N\r\}} \Omega \beta_\text{Y}^F \nonumber \\
{}& +\l( p_{i,j-1,k,l}- p_{i,j,k,l} \r) \mathbf{1}_{\l\{l=N\r\}} \Omega \beta_\text{Y}^B \nonumber \\
{}& +\delta_X  \l[\l(i+1\r)p_{i+1,j,k,l}- i p_{i,j,k,l}\r] \nonumber\\
{}& +\delta_Y  \l[ \l(j+1\r)p_{i,j+1,k,l}- j p_{i,j,k,l} \r]\nonumber\\
{}& + \frac{\alpha}{ \Omega} \l[\l(i+1\r)\l(j+1\r)p_{i+1,j+1,k,l}-ijp_{i,j,k,l}   \r]\nonumber\\
{}& + \frac{\kappa_\text{X}}{ \Omega}  \l[\l(i+1\r) p_{i+1,j,k-1,l} - i \mathbf{1}_{\l\{k<N\r\}} p_{i,j,k,l}\r] \nonumber \\
{}& +\frac{\kappa_\text{Y}}{ \Omega}\l[\l(i+1\r) p_{i+1,j,k,l-1}- i \mathbf{1}_{\l\{l<N\r\}} p_{i,j,k,l} \r]\nonumber\\
{}& +\theta_\text{X}  \l[p_{i-1,j,k+1,l} -  \mathbf{1}_{\l\{k>0\r\}} p_{i,j,k,l}\r] \nonumber \\
{}&+\theta_\text{Y}\l[p_{i+1,j,k,l+1} - \mathbf{1}_{\l\{l>0\r\}}p_{i,j,k,l}\r]. \label{eq:CME-titration}
\end{align}
Here, $\mathbf{1}_{\l\{\text{condition}\r\}}$ is the characteristic function: it is equal to $1$ when the condition is true, otherwise $0$. We also impose the boundary conditions that $p_{i,j,k,l}=0$ if $i<0$ or $j<0$. 

The deterministic dynamics of the stochastic processes governed by Eqs.~\eqref{eq:CME-repressilator} and \eqref{eq:CME-titration} can be derived by performing the Kramers--Moyal expansion to the lowest-order expansion $\mathcal{O}\l(\Omega^0\r)$ \citep{vanKampen,gardiner,lin2019ScalingMethodsAccelerating}. The procedure yields a Liouville equation describing the corresponding deterministic dynamics in the infinite population limit $\Omega\rightarrow \infty$, in which regime the gene expression noise diminishes to zero.

\begin{algorithm}[h]
\caption{Evolutionary algorithm.} \label{alg:evo}
   \begin{algorithmic}
\Require Zeroth-generation parameter $\theta_0$, population size $M$, number of simulated generations $L$, selected fraction $\phi$, mutation kernel $\mu\l(\theta'\vert \theta\r)$.
\newcommand{\algrule}[1][.2pt]{\par\vskip.3\baselineskip\hrule height #1\par\vskip.5\baselineskip}
\algrule
\State{$\ell \gets 0$} \Comment{Initialize}
\For{$i$ in $\l\{1,\ldots M\r\}$} \Comment{Initialize first generation}
    \State{$\theta_0^{[i]} \gets \theta \sim \mu \l( \cdot \vert \theta_0\r) $} \Comment{Assign randomized $\theta$ to $i^\text{th}$ individual}
\EndFor
\For{$\ell$ in $\l\{1,\ldots L\r\}$} 
    \For{$i$ in $\l\{1,\ldots M\r\}$}
        \State{Simulate time series $\l\{x(t_i)\r\}_i$ with $\theta_{\ell}^{[i]}$} 
        \State{Compute $\text{PSD}\l(f\r)$, $\text{PSD}_{\max}$, $f_{\max}$, and evaluate fitness $F\l(\theta^{[i]}_\ell \r)$}
    \EndFor
    \State{$S_\ell \gets \l\{\theta_{\ell}^{[i]} {\big\vert} F\l(\theta^{[i]}_\ell \r) \text{ ranks in top } \phi \text{ fraction of }M \r\}$}    \Comment{Selection}
    \For{$i$ in $\l\{1,\ldots M\r\}$} \Comment{Reproduction}
        \State{$u \sim \text{Uniform}(\l\{1, 2, \ldots \phi M\r\})$} \Comment{Choose one to reproduce}
        \State{$\theta_{\ell+1}^{[i]} \gets \theta\sim \mu\l(\cdot \vert \theta_{\ell}^{[u]}\r) $} \Comment{Reproduce an offspring with mutated $\theta$}
    \EndFor
\EndFor
\end{algorithmic}
\end{algorithm}

\beginsupplement

\end{document}